\documentclass[10pt,a4paper]{book}
\usepackage{udineHyderabad-Klu2}

\begin{document}
%%%%%%%%%%%%%%%%%%%%%%%%%%%%%%%%%%%%%%%%%%%%%%%%%%%%%%%%%%%%%%%%%%%%%%%%%%%%%%%%%%%%

	\pagestyle{fancy}
	
\talktitle{Data mining in gamma astrophysics experiments}{Data Mining in Gamma Astrophysics Experiments}

\talkauthors{Marco~Frailis\structure{1}, 
             Alessandro~De~Angelis\structure{1},
             Vito~Roberto\structure{2}}

\authorstucture[1]{INFN and Dipartimento di Fisica,
                   Università degli Studi di Udine, Italy}

\authorstucture[2]{Dipartimento di Matematica e Informatica,
                   Università degli Studi di Udine, Italy}

\shorttitle{Data mining in gamma astrophysics experiments}

\firstauthor{M.~Frailis et al.}

		\index{Frailis@\textsc{Frailis}, M.}%arco}
		\index{De Angelis@\textsc{De Angelis}, A.}%lessandro}
		\index{Roberto@\textsc{Roberto}, V.}%ito}

\begin{abstract}
Data mining techniques, including clustering and classification tasks, for the automatic information extraction from large datasets are increasingly demanded in several scientific fields. In particular, in the astrophysical field, large archives and digital sky surveys with dimensions of $10^{12}$~bytes currently exist, while in the near future they will reach sizes of the order of~$10^{15}$. In this work we propose a multidimensional indexing method to efficiently query and mine large astrophysical datasets. A novelty detection algorithm, based on the Support Vector Clustering and using density and neighborhood information stored in the index structure, is proposed to find regions of interest in data characterized by isotropic noise. We show an application of this method for the detection of point sources from a gamma-ray photon list.
\end{abstract}

%=============================================================
\section{Characterization of the astrophysical datasets}
%=============================================================
At present, several projects for the multi-wavelength observation of the Universe are underway, for example~SDSS, GALEX, POSS2, DENIS,~etc. In the next years, new spatial missions will be launched (e.g.~GLAST, Swift), surveying the wall sky at different wavelengths (gamma-ray, X-ray, optical). 
In the Astroparticle and Astrophysical fields, data is mostly characterized by multidimensional arrays. For instance, in X-ray and Gamma-ray astronomy, the data gathered by detectors are lists of detected photons whose properties include position (RA, DEC), arrival time, energy, error measures both for the position and the energy estimates, quality measures of the events . Source catalogs, produced by the analysis of the raw data, are lists of point and extended sources characterized by coordinates, magnitude, spectral indexes, flux,~etc. 

%=============================================================
\subsection{Mining multidimensional data}
%=============================================================
Data mining applied to multidimensional data analyzes the relationships between the attributes of a multidimensional object stored into the database and the attributes of the neighboring ones. Several queries are required by this kind of analysis:
\begin{itemize}
\item \emph{point queries}, to find all objects overlapping the query point;
\item \emph{range queries}, to find all objects having at least one common point with a query window;
\item \emph{nearest-neighbor queries}, to find all objects that have a minimum distance from the query object.
\end{itemize}
Another important operation is the \emph{spatial join}, which in the astrophysical field is needed to search multiple source catalogs and cross-identify sources from different wavebands. This multidimensional (spatial) data tend to be large (sky maps can reach sizes of Terabytes) requiring the integration of the secondary storage, and there is no total ordering on spatial objects preserving spatial proximity. This characteristic makes it di cult to use traditional indexing methods, like B+-trees or linear hashing.

%=============================================================
\section{An optimized R-tree structure}
%=============================================================
The {\bf R-tree} is a data structure meant to efficiently index multidimensional point data or objects with a spatial extent. The structure of an R-tree is the following:
\begin{itemize}
\item an {\bf inner node} of the R-tree has entries of the form (cp, MBB), where cp is the address of a child node and MBB is the n-dimensional Minimum Bounding Box of all entries in that child node;
\item a {\bf leaf node} has entries of the form (\emph{cp, MBB}), where \emph{cp} refers to a record describing a particular object and \emph{MBB} is its minimum bounding box, or (\emph{Point, Attributes}), where \emph{Point} is a coordinate in the n-dimensional space and \emph{Attributes} are data associated to that point.
\end{itemize}
An optimized index, in terms of construction time, memory occupied and query performances, can be built using a priori information on the dataset by means of bulk loading algorithms. We have followed a top-down construction method called VAMSplit algorithm to build and optimized R-Tree. This method preserves the spatial proximity between sibling nodes, resulting in a partition of the dataset with no overlapping between MBBs. Moreover, the volume of the data space covered by each node (at a particular level) is variable and dependent on data density. The main idea of this method is to recursively split the dataset on a near median element along the dimension with maximum variance. In particular, at each recursive step, child subtrees capacity is calculated by:
$$
cscap = B \cdot F ^{\big\lceil \log_{F} \big\lceil \frac{N}{B} \big\rceil \big\rceil - 1}
$$
where $B$ is the page capacity, $F$ is the internal fanout and $N$~is the number of elements to index in the current step. The near median element is computed by:
$$
med = cscap \cdot \Bigg\lfloor \frac{1}{2} \cdot \bigg\lceil \frac{N}{cscap} \bigg\rceil \Bigg\rfloor
$$
Our implementation uses a sampling strategy to find a good pivot value in the partition step and reduce the number of I/O operations; a caching strategy has been adopted to partition the data into the secondary storage. The total construction time is 
$O \big( \frac{N}{B}\log_{\frac{M}{B}}\frac{N}{B} \big)$.
\begin{figure}[tb]
\centerline{\includegraphics[width=\textwidth]{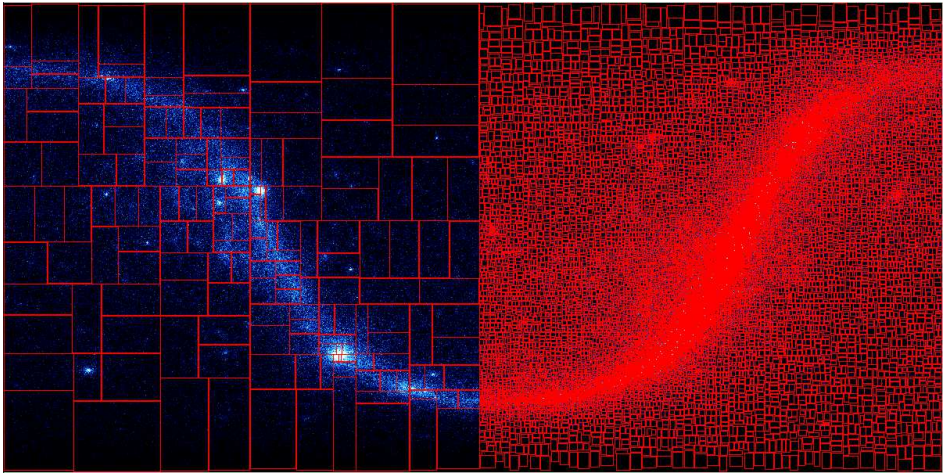}}
\caption{The structure\label{fig:R-tree} of the optimized R-tree built a photon dataset}
\end{figure}

%=============================================================
\section{A scalable novelty detection algorithm}
%=============================================================
The structure of the optimized R-tree can help exploring the data and finding regions of interest. For this purpose, other information can be added to each node: the total number of data points covered by the node, their mean and variance, other statistical moments. In this work we propose a novelty detection algorithm based on the \emph{Support Vector Clustering}~(SVC).

%=============================================================
\subsection{The SVC algorithm}
%=============================================================
The SVC algorithm estimates the \emph{support} of a high dimensional distribution.
It computes the hypersphere with minimal radius which encloses the data points when mapped to a high dimensional feature space.
Given a set of points $\mathbf{x}_1,\ldots,\mathbf{x}_N$\hspace*{-.5pt}, with $\mathbf{x}_i \in X \subseteq \mathbb{R}^d$, it finds the hypersphere $(\mathbf{c},r)$ that solves the optimization problem:
$$
\begin{array}{ll}
\begin{aligned}
 \min_{\mathbf{c},r,\rho}
\ & \ 
 r^2 + C \sum_{i=1}^{N} \rho_i
\\[2mm]
 \textrm{s.t.}     
\ & \ 
 \| \phi(\mathbf{x}_i) - \mathbf{c} \| \leq r^2 + \rho_i
\\[2mm]                  
& \ 
 \rho_i \geq 0, \quad i=1,\ldots,N
\end{aligned}  
\end{array}
$$
where $\rho_i$ are slack variables and $C$ is a positive constant.
Defining the Lagrangian, the solution is obtained solving the dual problem:
$$
\begin{array}{ll}
\begin{aligned}
 \max_{\mathbf{\alpha}}
\ & \ 
 \sum_{i=1}^{N}   \alpha_i         k(\mathbf{x}_i,\mathbf{x}_i) - 
 \sum_{i,j=1}^{N} \alpha_i\alpha_j k(\mathbf{x}_i,\mathbf{x}_j)
\\[2mm]
 \textrm{s.t.}     
\ & \ 
 0 \leq \alpha_i \leq C, \quad i=1,\ldots,N
\end{aligned}  
\end{array}
$$
where $\alpha_i$ are the lagrange multipliers and $k(\mathbf{x}_i,\mathbf{x}_j) = \langle\phi(\mathbf{x}_i),\phi(\mathbf{x}_j)\rangle$.

%=============================================================
\subsection{Features from the R-tree}
%=============================================================
The partition generated at a given level of the optimized R-tree is used as the input space of the novelty detection algorithm. For each node of the partition, the input parameters include the \emph{center~\bf{c}} of its bounding box and the \emph{density} (the ratio between the number~$n$ of elements covered by the node and the volume~$V$ of its bounding box). 
These features are not orthogonal.
Therefore, we first apply the PCA method to find the directions along which the variance is higher and project the features on the corresponding eigenvectors.
The projected features are passed to the SVC algorithm.

%=============================================================
\subsection{Gamma source detection}
\begin{figure}[tb]
\begin{center}
\includegraphics[height=31mm,width=0.495\textwidth]{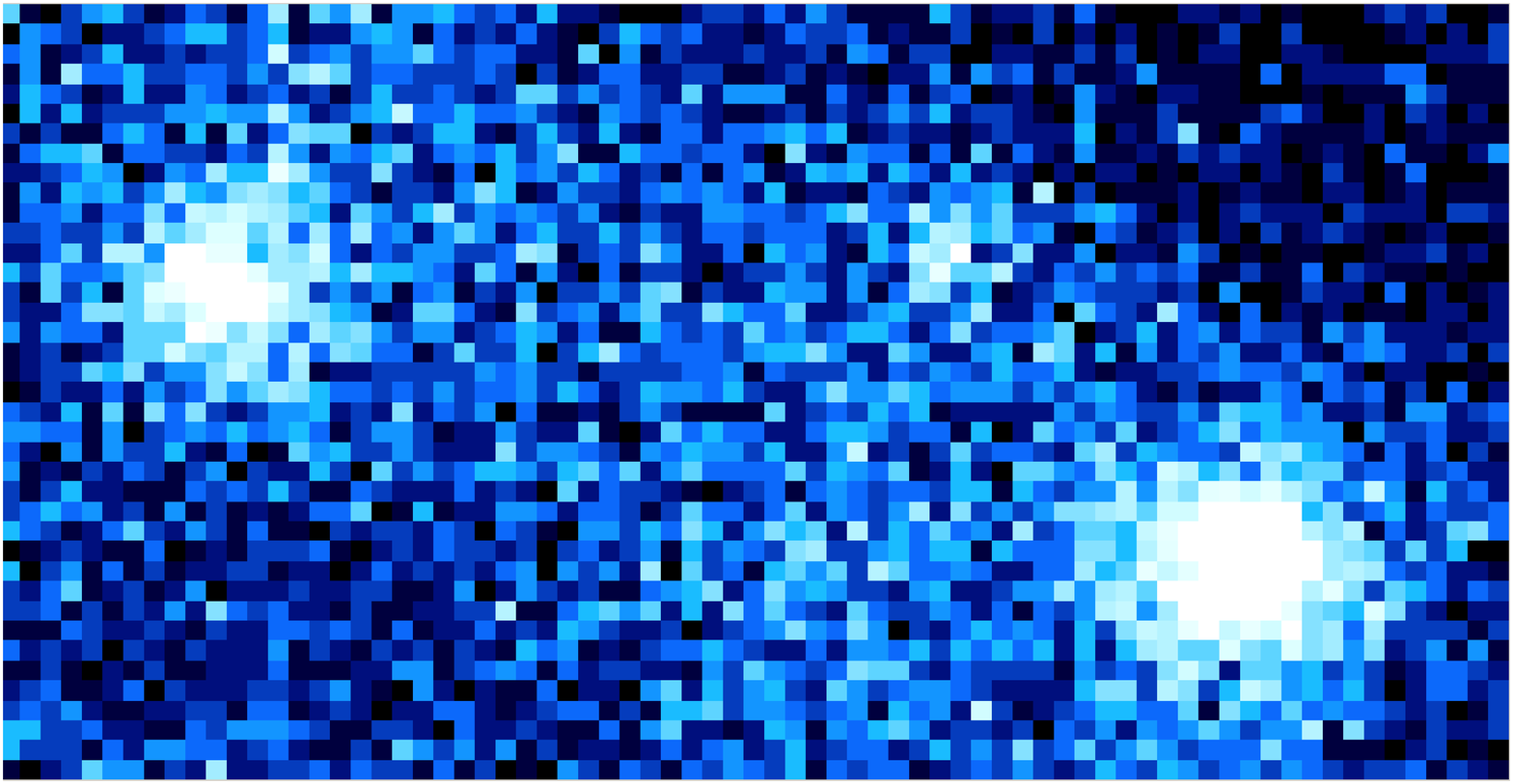}
\includegraphics[width=0.495\textwidth]{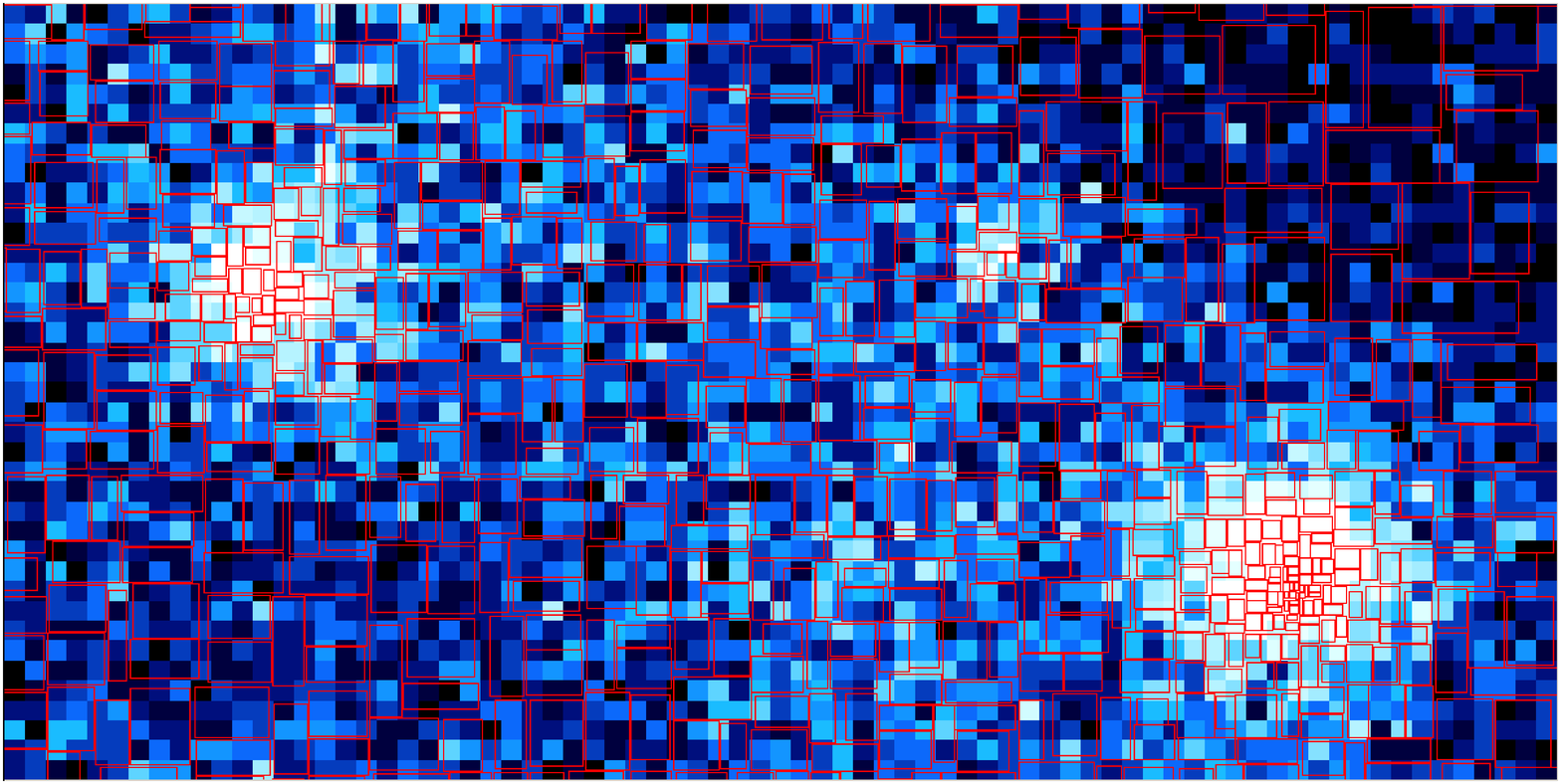}
\\[1mm]
\includegraphics[width=0.495\textwidth]{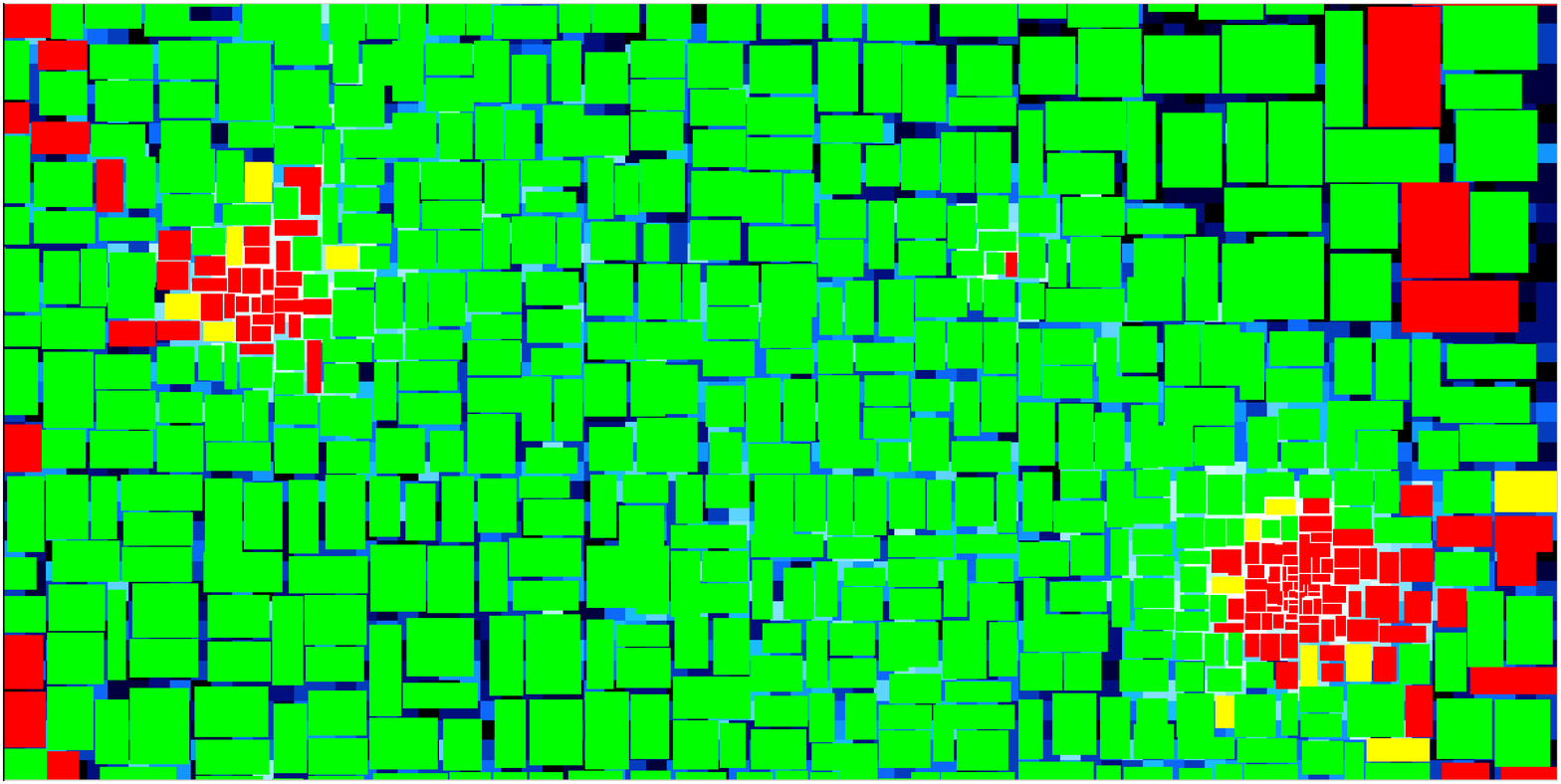}
\includegraphics[width=0.495\textwidth]{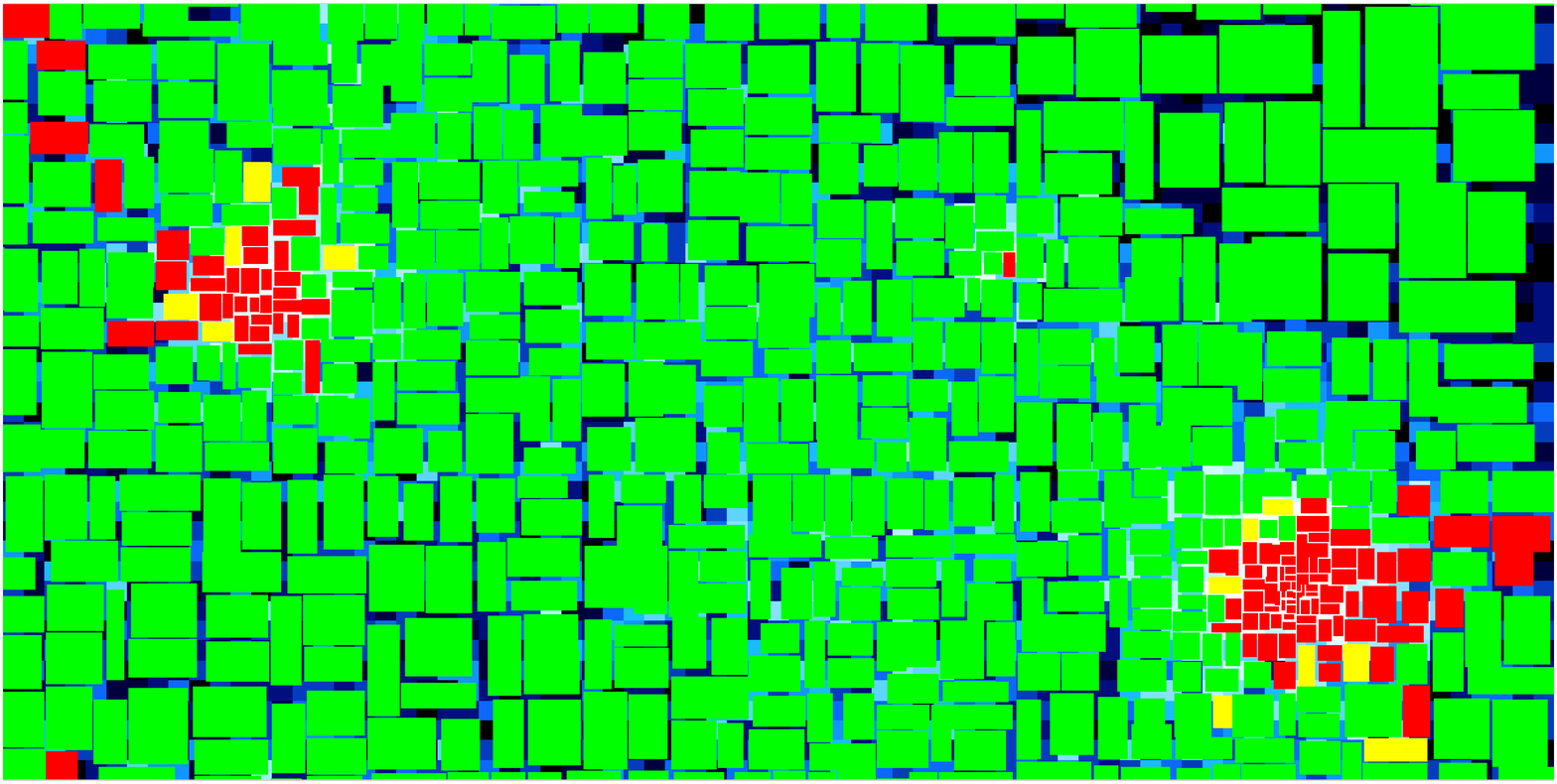}
\end{center}
\caption{The novelty detection algorithm\label{fig:novelty} applied to the anticenter}
\end{figure}

%=============================================================
Point sources are mostly characterized by a stronger flux, with respect to the surrounding, focused on a small angular region. The area covered by a point source depends also on the instrument point spread function. 
An optimized R-tree index can be built on a dataset including photons gathered in a certain range of time (we are using, for the analysis, a minimum interval of 6 days). To find static or strong variable sources (e.g.~gamma-ray bursts) we use only a bidimensional indexing on the RA and DEC~values.
The algorithm estimates the support of the diffuse background.
The output of the SVC algorithm is filtered to single out the nodes with lowest density. Point sources are considered as the remaining outliers.
Figure~\ref{fig:novelty} shows the application of our method to the anticenter region. Green boxes represent the background (support) while yellow boxes are \emph{support vectors} and the red ones are the outliers. In particular,the three major sources of the anticenter are highlighted as novelty.

%=============================================================
\section{Conclusions}
%=============================================================
In this work we have realized a multidimensional indexing method to efficiently access and mine large multidimensional astrophysical data. The index adapts the VAMSRtree to large datasets. The partition generated by  the optimized R-tree is used to scale the SVC algorithm and find regions of interest where a more accurate analysis can be performed.

%=============================================================

%=============================================================

%%%%%%%%%%%%%%%%%%%%%%%%%%%%%%%%%%%%%%%%%%%%%%%%%%%%%%%%%%%%%%%%%%%%%%%%%%%%%%%%%%%%
\end{document}